\documentclass[twocolumn,showpacs,aps,prl,superscriptaddress,floatfix]{revtex4}

\usepackage{graphicx}
\usepackage{dcolumn}
\usepackage{amsmath}
\usepackage{epsfig}

\def\sss{\scriptscriptstyle}
\def\barpd{{\raise.35ex\hbox
{${\sss (}$}}--{\raise.35ex\hbox{${\sss )}$}}}
 \def\dbarp{\tilde{D}^0}
 \def\dbarpnozero{\tilde{D}}

\def\Dz      {\ensuremath{D^0}\xspace}

\def\mes        {\mbox{$m_{\rm ES}$}\xspace}

\def\to         {\ensuremath{\rightarrow}\xspace}

\newcommand{\BaBarYear}       {04}
\newcommand{\BaBarNumber}     {002}
\newcommand{\SLACPubNumber} {10333}

\RequirePackage{xspace}

\usepackage{relsize}
\def\babar{\mbox{\slshape B\kern-0.1em{\smaller A}\kern-0.1em
    B\kern-0.1em{\smaller A\kern-0.2em R}}}

\def\Kbar  {\kern 0.2em\overline{\kern -0.2em K}{}\xspace}

\def\Kz    {\ensuremath{K^0}\xspace}
\def\Kzb   {\ensuremath{\Kbar^0}\xspace}
\def\KzKzb {\ensuremath{\Kz \kern -0.16em \Kzb}\xspace}
\def\Kp    {\ensuremath{K^+}\xspace}
\def\Km    {\ensuremath{K^-}\xspace}

\def\KpKm  {\ensuremath{\Kp \kern -0.16em \Km}\xspace}

\def\Dbar    {\kern 0.2em\overline{\kern -0.2em D}{}\xspace}

\def\Dz      {\ensuremath{D^0}\xspace}
\def\Dzb     {\ensuremath{\Dbar^0}\xspace}
\def\DzDzb   {\ensuremath{\Dz {\kern -0.16em \Dzb}}\xspace}
\def\Dp      {\ensuremath{D^+}\xspace}
\def\Dm      {\ensuremath{D^-}\xspace}

\def\DpDm    {\ensuremath{\Dp {\kern -0.16em \Dm}}\xspace}

\def\Bbar    {\kern 0.18em\overline{\kern -0.18em B}{}\xspace}

\def\Bz      {\ensuremath{B^0}\xspace}
\def\Bzb     {\ensuremath{\Bbar^0}\xspace}
\def\BzBzb   {\ensuremath{\Bz {\kern -0.16em \Bzb}}\xspace}
\def\Bu      {\ensuremath{B^+}\xspace}
\def\Bub     {\ensuremath{B^-}\xspace}

\def\BpBm    {\ensuremath{\Bu {\kern -0.16em \Bub}}\xspace}

\def\BorBbar    {\kern 0.18em\optbar{\kern -0.18em B}{}\xspace}
\def\DorDbar    {\kern 0.18em\optbar{\kern -0.18em D}{}\xspace}
\def\KorKbar    {\kern 0.18em\optbar{\kern -0.18em K}{}\xspace}

\mathchardef\Upsilon="7107
\def\Y#1S{\ensuremath{\Upsilon{(#1S)}}\xspace}

\def\FourS {\Y4S}

\mathchardef\Deltares="7101
\mathchardef\Xi="7104
\mathchardef\Lambda="7103
\mathchardef\Sigma="7106
\mathchardef\Omega="710A

\def\Deltabar{\kern 0.25em\overline{\kern -0.25em \Deltares}{}\xspace}
\def\Lbar{\kern 0.2em\overline{\kern -0.2em\Lambda\kern 0.05em}\kern-0.05em{}\xspace}
\def\Sigbar{\kern 0.2em\overline{\kern -0.2em \Sigma}{}\xspace}
\def\Xibar{\kern 0.2em\overline{\kern -0.2em \Xi}{}\xspace}
\def\Obar{\kern 0.2em\overline{\kern -0.2em \Omega}{}\xspace}
\def\Nbar{\kern 0.2em\overline{\kern -0.2em N}{}\xspace}
\def\Xb{\kern 0.2em\overline{\kern -0.2em X}{}\xspace}

\def\mes        {\mbox{$m_{\rm ES}$}\xspace}

\newcommand{\tev}{\ensuremath{\mathrm{\,Te\kern -0.1em V}}\xspace}
\newcommand{\gev}{\ensuremath{\mathrm{\,Ge\kern -0.1em V}}\xspace}
\newcommand{\mev}{\ensuremath{\mathrm{\,Me\kern -0.1em V}}\xspace}
\newcommand{\kev}{\ensuremath{\mathrm{\,ke\kern -0.1em V}}\xspace}
\newcommand{\ev}{\ensuremath{\mathrm{\,e\kern -0.1em V}}\xspace}
\newcommand{\gevc}{\ensuremath{{\mathrm{\,Ge\kern -0.1em V\!/}c}}\xspace}
\newcommand{\mevc}{\ensuremath{{\mathrm{\,Me\kern -0.1em V\!/}c}}\xspace}
\newcommand{\gevcc}{\ensuremath{{\mathrm{\,Ge\kern -0.1em V\!/}c^2}}\xspace}
\newcommand{\mevcc}{\ensuremath{{\mathrm{\,Me\kern -0.1em V\!/}c^2}}\xspace}

\def\mus  {\ensuremath{\rm \,\mus}\xspace}

\def\mus        {\ensuremath{\,\mu{\rm s}}\xspace}    

\def\to                 {\ensuremath{\rightarrow}\xspace}

\def\pep2{PEP-II}
\def\BF{$B$ Factory}

\def\gsim{{~\raise.15em\hbox{$>$}\kern-.85em
          \lower.35em\hbox{$\sim$}~}\xspace}
\def\lsim{{~\raise.15em\hbox{$<$}\kern-.85em
          \lower.35em\hbox{$\sim$}~}\xspace}

\def\CP                {\ensuremath{C\!P}\xspace}

\xspace

\def\jetset74   {\mbox{\tt Jetset \hspace{-0.5em}7.\hspace{-0.2em}4}\xspace}

\begin{document}

   {\pagestyle{empty}
   \begin{flushleft}
   \babar-PUB-\BaBarYear/\BaBarNumber \\
   SLAC-PUB-\SLACPubNumber \\
   {\tt arXiv:hep-ex/0402024} \\
   Phys. Rev. Lett. {\bf 93}, 131804 (2004) \\
   \end{flushleft}
   }

 \title{Search for {\boldmath $B^{\pm} \to [K^{\mp}\pi^{\pm}]_D K^{\pm}$}
 and upper limit on the {\boldmath $b\rightarrow u$ amplitude in $B^\pm \to D K^\pm$}}

%
\author{B.~Aubert}
\author{R.~Barate}
\author{D.~Boutigny}
\author{F.~Couderc}
\author{J.-M.~Gaillard}
\author{A.~Hicheur}
\author{Y.~Karyotakis}
\author{J.~P.~Lees}
\author{V.~Tisserand}
\author{A.~Zghiche}
\affiliation{Laboratoire de Physique des Particules, F-74941 Annecy-le-Vieux, France }
% ...
\collaboration{The \babar\ Collaboration}
\noaffiliation

\date{\today}

\begin{abstract}
\noindent
We search for $B^{\pm} \to [K^{\mp}\pi^{\pm}]_D K^{\pm}$ decays, where 
$[K^{\mp}\pi^{\pm}]_D$ indicates that the $K^{\mp}\pi^{\pm}$ pair
originates from the decay of a $D^0$ or \Dzb.  Results are based on
$120 \times 10^6$ $\FourS \to B\Bbar$ decays collected with the
\babar\ detector at SLAC.  We set an upper limit on the ratio
    \[
    {\cal R}_{K\pi} \equiv 
    \frac{(\Gamma(B^+ \to [K^-\pi^+]_D K^+) +  \Gamma(B^- \to [K^+\pi^-]_D K^-)) }
         {(\Gamma(B^+ \to [K^+\pi^-]_D K^+) +  \Gamma(B^- \to [K^-\pi^+]_D K^-)) } <  0.026
         \ \ {\rm (90\% \ C.L.).}
    \]
This constrains the amplitude ratio 
$r_B \equiv |A(B^- \to \Dzb K^-)/A(B^- \to D^0 K^-)| < 0.22$ (90\% C.L.),
consistent with expectations.  The small value of $r_B$ favored by our 
analysis suggests that the determination of the CKM phase $\gamma$
from
  $B\rightarrow D K$
will be difficult.

\end{abstract}


\maketitle


   Following the discovery of \CP violation in $B$-meson 
   decays and the measurement of the angle $\beta$
   of the unitarity triangle~\cite{cpv} associated with
   the Cabibbo-Kobayashi-Maskawa (CKM) quark mixing matrix, focus has turned
   towards the measurements of the other angles $\alpha$ and $\gamma$.
   The angle $\gamma$ is ${\rm arg}(-V_{ub}^*V^{}_{ud}/V_{cb}^*V^{}_{cd})$,
   where $V^{}_{ij}$ are CKM matrix elements;
   in the Wolfenstein convention~\cite{wolfenstein},
   $\gamma = {\rm arg}(V_{ub}^*)$.

\begin{figure}[hb]
      \epsfig{file=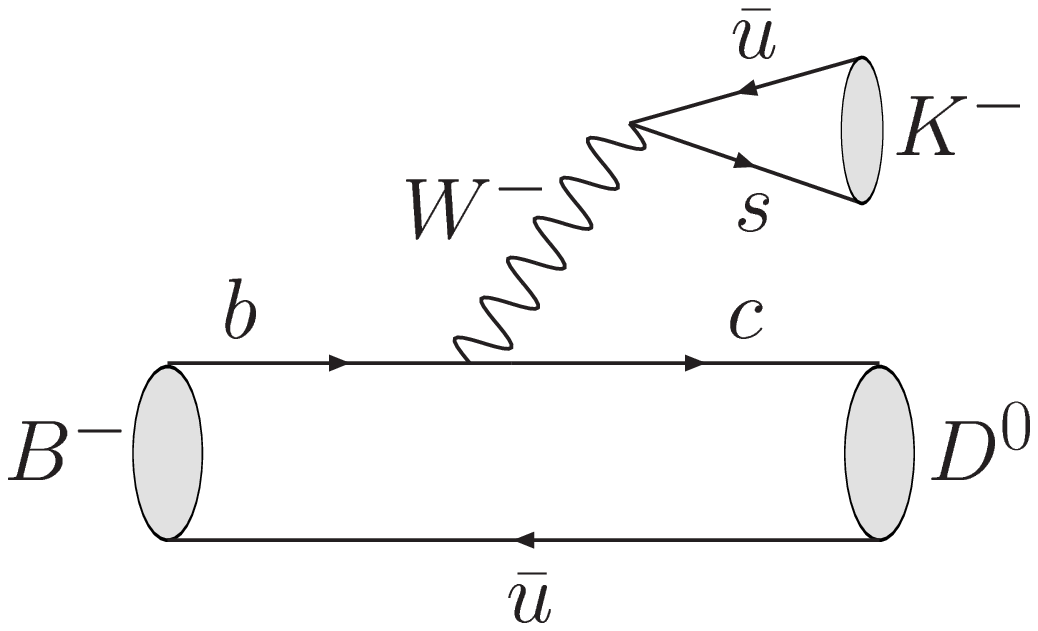,height=2.9cm}
      \epsfig{file=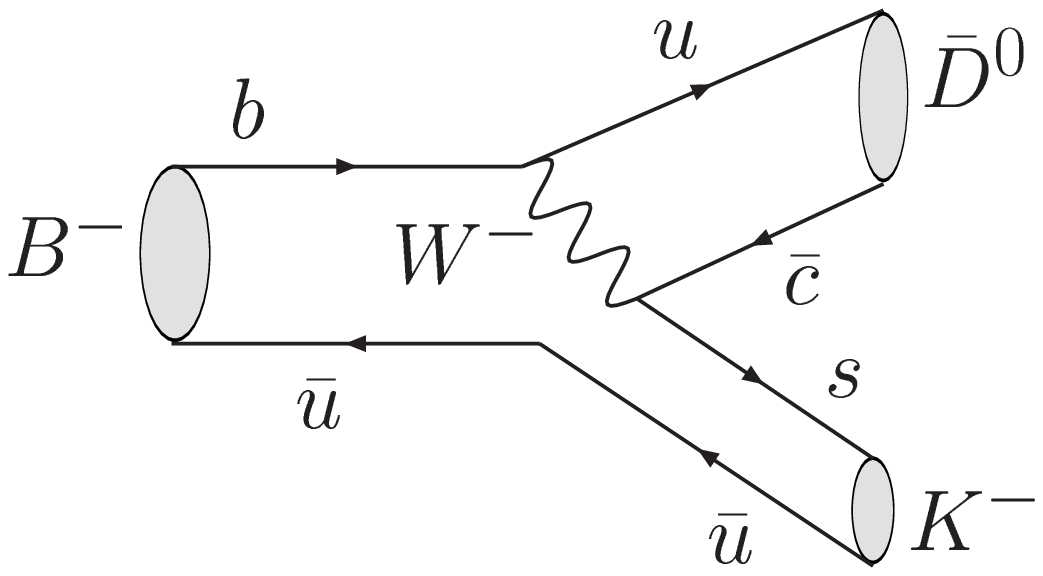,height=2.9cm}
   \caption{Feynman diagrams for $B^- \to D^0 K^-$ and $\Dzb K^-$.
   The latter is CKM- and color-suppressed with respect to the former.}
   \label{fig:feynman}
   \end{figure}

   Several proposed methods for measuring $\gamma$ exploit the
   interference between $B^- \to D^0 K^-$ and $B^- \to \Dzb K^-$
   (Fig.~\ref{fig:feynman}) which occurs when the $D^0$ and the
   \Dzb decay to common final states, as first suggested in Ref.~\cite{dk1}.

   Following the proposal in Ref.~\cite{dk2}, we search for
   $B^- \to \dbarp~K^-$
   followed by
   $\dbarp~\to K^+\pi^-$,
   as well as the charge conjugate
   sequence,
   where the symbol $\dbarp$ indicates either a $D^0$ or a
   $\overline{D}^0$.
   Here the favored $B$ decay followed by the doubly 
   CKM-suppressed $D$ decay interferes with the suppressed $B$ decay followed
   by the CKM-favored $D$ decay.  We use the notation $B^- \to [h_1^+h_2^-]_D h_3^-$
   (with each $h_i=\pi$ or $K$) for the decay chain $B^- \to \dbarp~h_3^-$,
   $\dbarp~\to h_1^+ h_2^-$.  We also refer to $h_3$ as the bachelor
   $\pi$ or $K$.  Then, ignoring $D$ mixing,

   \begin{equation}
   {\cal R}^{\pm}_{K\pi} \equiv 
   \frac{\Gamma([K^\mp \pi^\pm]_D K^\pm)}{\Gamma([K^\pm \pi^\mp]_D K^\pm)}
   = r_B^2 + r_D^2 + 2 r_B r_D \cos(\pm \gamma + \delta),
   \nonumber
   \end{equation}
   \noindent where
   \begin{equation}
   r_B \equiv \left| \frac{A(B^- \to \Dzb K^-)}{A(B^- \to D^0 K^-)} \right|, 
   \ \ \delta \equiv \delta_B + \delta_D,
   \nonumber
   \end{equation}
   \begin{equation}
   r_D \equiv \left| \frac{A(D^0 \to K^+ \pi^-)}{A(D^0 \to K^- \pi^+)} \right|
   = 0.060 \pm 0.003~\mbox{\cite{dcsdbr}},
   \nonumber
   \end{equation}
   and \noindent $\delta_B$ and $\delta_D$ are 
   strong phase differences between the two $B$ and $D$ decay 
   amplitudes, respectively.  The expression for 
   ${\cal R}^\pm_{K\pi}$ neglects the tiny contribution to
   the $[K^\pm \pi^\mp]_D K^\pm$ mode from the color suppressed
   $B$-decay followed by the douby-CKM suppressed $D$-decay.

   Since $r_B$ is expected to be of the same order as $r_D$,
   \CP violation could manifest itself as a large difference between
   ${\cal R}^+_{K\pi}$ and ${\cal R}^-_{K\pi}$.  Measurements of
   ${\cal R}^\pm_{K\pi}$ are not sufficient to extract $\gamma$,
   since these two quantities are functions of three
   unknowns: $\gamma$, $r_B$, and $\delta$.  However, they can be combined 
   with measurements for other $\dbarp$ modes to extract $\gamma$
   in a theoretically clean way~\cite{dk2}.

   The value of $r_B$ determines, in part, the level of interference
   between the diagrams
   of Fig.~\ref{fig:feynman}.  In most techniques for measuring $\gamma$,
   high values of $r_B$ lead to 
   better sensitivity.
   Since ${\cal R}^\pm_{K\pi}$ depend quadratically on $r_B$, measurements
   of ${\cal R}^\pm_{K\pi}$ can constrain $r_B$.
   In the Standard Model,
   $r_B = |V^{}_{ub} V_{cs}^*/V^{}_{cb} V_{us}^*| \: F_{cs} \approx 0.4 \: F_{cs}$,
   and $F_{cs} < 1$ accounts for the additional suppression, beyond that
   due to CKM factors, of $B^- \to \Dzb K^-$ relative to $B^- \to D^0 K^-$.
   Naively, $F_{cs} = \frac{1}{3}$, which is the probability for the color
   of the quarks from the virtual $W$ in $B^- \to \Dzb K^-$ to match 
   that of the other two quarks; see Fig.~\ref{fig:feynman}.  Early estimates 
   gave $F_{cs} \approx 0.22$~\cite{neubert}, leading to
   $r_B \approx 0.09$; however, recent measurements~\cite{colorsuppressed} 
   of color suppressed $b \to c$ decays ($B \to D^{(*)} h^0$; 
   $h^0 = \pi^0, \rho^0, \omega, \eta, \eta'$) suggest that $F_{cs}$, and
   therefore $r_{B}$, could be larger, {\em e.g.}, 
   $r_B \approx 0.2$~\cite{gronaurb}.
   A study by the Belle collaboration
   of $B^{\pm} \to \dbarp K^{\pm}$, $\dbarp \to K_S \pi^+ \pi^-$,
   favors a large value of $r_B$: $r_B = 0.26^{+0.11}_{-0.15}$~\cite{belleRb}.

   Our results are based on $120 \times 10^6$ $\FourS\to B\Bbar$ decays,
   corresponding to an integrated luminosity of 109 fb$^{-1}$, collected
   between 1999 and 2003 with the \babar\ detector~\cite{babar} at the \pep2\
   \BF\ at SLAC.  
   A 12~fb$^{-1}$ off-resonance data sample,
   with a CM energy 40~\mev below the \FourS resonance,
   is used to study continuum events, $e^+ e^- \to q \bar{q}$
   ($q=u,d,s,$ or $c$).

   The event selection was developed from studies of
   simulated $B\Bbar$ and continuum events, and off-resonance
   data. A large on-resonance data sample of $B^- \to D^0 \pi^-$,
   $D^0 \to K^- \pi^+$ events was used to validate several aspects of the
   simulation and analysis procedure. We refer to this mode
   and its charge conjugate as $B \to D\pi$.

   Kaon and pion candidates in $B^{\pm} \to [K \pi]_D K^{\pm}$
   must satisfy $K$ or $\pi$ identification 
   criteria
   that are typically 90\% efficient, depending on momentum and polar angle.
   Misidentification rates are at the few percent level.
   The invariant mass of the $K\pi$ pair must be within 18.8 MeV
   (2.5$\sigma$) of the mean reconstructed $D^0$ mass.
   The remaining background from other $B^\pm \to [h_1 h_2]_D h_3^\pm$ modes
   is eliminated by removing events where any $h_i^{+} h_j^{-}$
   pair, with any particle-type assignment except for the signal hypothesis
   for the $h_1 h_2$ pair, is consistent with $\dbarp$ decay.
   We also reject $B$ candidates where the $\dbarp$ paired
   with a $\pi^0$ or $\pi^\pm$ in the  event is consistent
   with $D^* \to D \pi$ decay.

   After these 
   requirements, backgrounds are 
   mostly
   from continuum, 
   mainly 
   $e^+ e^- \to c \bar{c}$, with 
   $\bar{c} \to \Dzb \to K^+ \pi^-$ and $c \to D \to K^-$.
   These are reduced  
   with a neural network based
   on nine quantities that distinguish continuum and $B\Bbar$ events:  
   (i) A Fisher discriminant
   based on  
   the quantities $L_0 = \sum_i{p_i}$ and $L_2 = \sum_i{p_i \cos^2\theta_i}$
   calculated in the CM frame.  Here, $p_i$ is the momentum and
   $\theta_i$ is the angle with respect to the thrust axis of the $B$ candidate
   of tracks and clusters not used to reconstruct the $B$.
   (ii) $|\cos \theta_T|$, where $\theta_T$ is the angle in
   the CM frame between the thrust axes of the $B$ 
   and the detected remainder of the event.
   (iii) $\cos \theta_B$, where $\theta_B$ is the polar angle
   of the $B$ 
   in the CM frame.
   (iv) $\cos \theta_D^K$ where $\theta_D^K$ is the decay angle
   in $\dbarp \to K\pi$, {\it i.e.}, the angle between the direction of the $K$ 
   and the line of flight of the $\dbarp$ in the $\dbarp$ rest frame.
   (v) $\cos \theta_B^D$, where $\theta_B^D$ is the decay angle
   in $B \to \dbarp K$.
   (vi) the 
   difference $\Delta Q$ between the sum
   of the charges of tracks in the $\dbarp$ hemisphere
   and the sum of the charges of the tracks in the opposite
   hemisphere excluding the tracks used in the reconstructed $B$.
   For signal, $\langle \Delta Q \rangle = 0$, 
   while
   for the $c\bar{c}$ background 
   $\langle \Delta Q \rangle \approx \frac{7}{3}\times Q_B$,
   where $Q_B$ is the $B$ candidate charge.  The
   $\Delta Q$ RMS is 2.4.
   (vii) $Q_B \cdot Q_K$, where $Q_K$ is the sum of the charges of all
   kaons not in the reconstructed $B$.
   Many signal events have $Q_B \cdot Q_K \leq -1$,
   while most continuum events have no kaons outside of the
   reconstructed $B$, and 
   hence
   $Q_K = 0$.
   (viii) the distance of closest approach between the bachelor
   track and the trajectory of the $\dbarp$.  This is
   consistent with zero for signal events, but can be
   larger in $c\bar{c}$ events.
   (ix) the existence of a lepton ($e$ or $\mu$) and
   the invariant mass ($m_{K\ell}$) of 
   the
   lepton and the bachelor $K$.
   Continuum events have fewer leptons than signal events. 
   Moreover, most
   leptons in $c\bar{c}$ events
   are from $D \to K \ell \nu$, where $K$ is the bachelor kaon,
   so that $m_{K\ell} < m_D$.

   The neural 
   net
   is trained with simulated 
   continuum and signal events.  
   We find agreement between the distributions of all nine variables
   in simulation and in control samples of off-resonance data
   and of $B \to D \pi$.  The neural net requirement is 66\% efficient
   for signal, and rejects 96\% of the continuum background.
   An additional requirement, $\cos\theta_D^K > -0.75$, rejects 
   50\%
   of the remaining $B\Bbar$ backgrounds and is 93\% efficient for signal.

   A $B$ candidate is characterized by the energy-substituted mass
   $\mes \equiv \sqrt{(\frac{s}{2}  + \vec{p}_0\cdot \vec{p}_B)^2/E_0^2 - p_B^2}$
   and energy difference $\Delta E \equiv E_B^*-\frac{1}{2}\sqrt{s}$, 
   where $E$ and $p$ are energy and momentum, the asterisk
   denotes the CM frame, the subscripts $0$ and $B$ refer to the
   \FourS and $B$ candidate, respectively, and $s$ is the square
   of the CM energy.  For signal events $\mes = m_B$ within the 
   resolution of about 2.5 MeV, where $m_B$ is the known $B$ mass.

   We require $\Delta E$ to be within 47.8 MeV ($2.5\sigma$)
   of the mean value of $-4.1$ MeV found in the $B \to D \pi$
   control sample.  The yield of signal events is extracted
   from a fit to the $\mes$ distribution of events satisfying
   all of the requirements discussed above.

   \begin{figure*}[tb]
      \includegraphics[width=\linewidth]{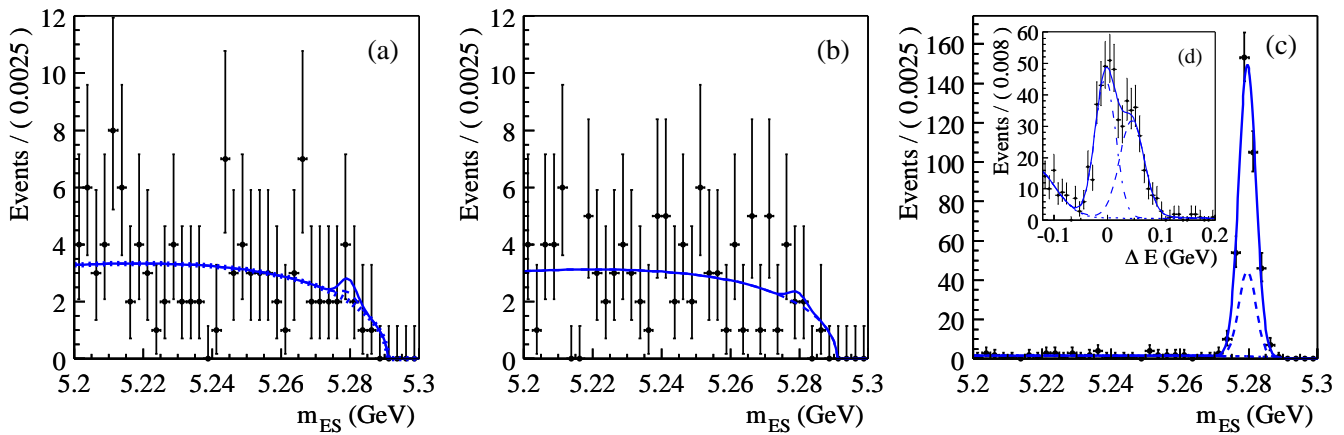}
   \caption{\mes distributions for (a) signal ($[K^\mp \pi^\pm]_D K^\pm$) 
   candidates, (b) candidates from the $\dbarp$ sideband, and (c) $B \to DK$ candidates.
   The $\dbarp$ sideband selection uses a $K^{\mp}\pi^{\pm}$ invariant
   mass range $2.72$ times larger than the signal selection.  (d) $\Delta E$
   distribution for $B \to DK$ candidates;  the peak
   centered at $\approx$ 0.05 GeV is from $B \to D \pi$. The superimposed 
   curves are described in the text.  In (c), the dashed Gaussian centered at
   $m_B$ represents the $B \to D\pi$ contribution estimated from (d).} 
   \label{fig:mes_de}
   \end{figure*}

   Our selection includes contributions from backgrounds
   with \mes distributions peaked near $m_B$ (peaking backgrounds).
   We distinguish those with a real $\dbarp \to K^{\mp}\pi^{\pm}$
   and those without, {\it e.g.}, $B^- \to h^+h^-h^-$.  The latter are 
   estimated 
   from 
   events with $K^{\mp}\pi^{\pm}$ mass in a sideband
   of the $\dbarp$.  The former are 
   from $B^- \to D^0 \pi^-$,
   followed by the CKM-suppressed decay $D^0 \to K^+\pi^-$, 
   with
   the bachelor
   $\pi$ 
   misidentified as a $K$.  These are estimated as
   $N_{peak}^D = r_D^2 N_{D\pi}$, where $N_{D\pi}$ is
   the number of observed $B \to D\pi$ events
   with the $\pi$ misidentified as a $K$.
   The technique used to measure $N_{D\pi}$ is described below.
   Studies of simulated $B\Bbar$ events indicate that
   other peaking background contributions are negligible.

   Because of the small number of events, we combine the 
   $B^{+}$ and $B^{-}$ samples.  We define the quantity 
   \begin{equation}
   {\cal R}_{K\pi} \equiv 
   \frac{\Gamma(B^- \to [K^+ \pi^-]_D K^-)+\Gamma(B^+ \to [K^- \pi^+]_D K^+)}
        {\Gamma(B^- \to [K^- \pi^+]_D K^-)+\Gamma(B^+ \to [K^+ \pi^-]_D K^+)},
   \nonumber
   \end{equation}
   \begin{equation}
   {\cal R}_{K\pi} = \frac{{\cal R}_{K\pi}^{+}+{\cal R}_{K\pi}^{-}}{2} =
   r_B^2 + r_D^2 + 2 r_B r_D \cos\gamma \cos \delta,
   \nonumber
   \end{equation}
   assuming no $CP$ violation in $[K^\mp \pi^\pm]_D K^\mp$.

   We determine ${\cal R}_{K\pi} = c N_{sig}/N_{DK}$, where $N_{sig}$ is
   the number of $B^\pm \to [K^{\mp} \pi^{\pm}]_D K^{\pm}$ signal events
   and $N_{DK}$ is the number of $B^{\pm} \to [K^{\pm} \pi^{\mp}]_D K^{\pm}$
   events, a mode which we denote by $B \to DK$.  Most systematic uncertainties
   cancel in the ratio.  The factor $c = 0.93 \pm 0.04$, determined from simulation, 
   accounts for a difference in the event selection efficiency between the
   signal mode and $B \to DK$.  This difference is mostly due to a correlation
   between the efficiencies of the $\cos\theta^K_D$ requirement and the $\dbarp$
   veto constructed using the bachelor track and the oppositely-charged track
   in the $[K\pi]$ pair.  This correlation depends on the relative sign
   of the kaon and the bachelor track, and is different in the two modes.

   The value of ${\cal R}_{K\pi}$ is obtained from a simultaneous unbinned
   maximum likelihood fit to four \mes and three $\Delta E$ distributions.
   These distributions are used to extract the parameters needed to calculate
   ${\cal R}_{K\pi}$ ({\em e.g.}, $N_{sig}$) or to constrain the shapes of other
   distributions.  The likelihood is expressed directly in terms of ${\cal R}_{K\pi}$.

   The $\mes$ distribution for signal candidates is fit to the sum of a 
   threshold background function 
   and a Gaussian centered at $m_B$.
   The number of events in the Gaussian is $N_{sig} + N_{peak}^D + N_{peak}^{hhh}$,
   where $N_{peak}^D$ and $N_{peak}^{hhh}$ are the number of peaking background events
   with and without a real $\dbarp$, respectively.  The Gaussian parameters
   are constrained by the fit to the $\mes$ distribution of $B \to DK$ events.
   The shape of the threshold function is constrained by fitting the $\mes$
   distribution of candidates in a sideband of $\Delta E$
   ($-125 < \Delta E < 200$~MeV, excluding the signal region).
   The $\mes$ distribution for events passing all signal requirements, but
   with $K^{\mp}\pi^{\pm}$ mass in the sideband of the $\dbarp$ is
   fit in the same manner.  We estimate $N_{peak}^{hhh}$ from the Gaussian
   yield of this last fit, accounting for the different sizes  of the
   signal and sideband $\dbarp$ mass ranges.  The $\mes$ distributions for signal and 
   $\dbarp$ sideband candidates are shown in Fig.~\ref{fig:mes_de}a,b.

   The $\mes$ distribution for $B \to DK$ candidates with
   $|\Delta E + 4.1 \ {\rm MeV}| < 47.8$ MeV (see Fig.~\ref{fig:mes_de}c)
   is also fit to a Gaussian and a threshold function.  The number of
   events in the Gaussian is $N_{DK} + N_{D\pi}$, where, as previously defined,
   $N_{DK}$ is the number of $B \to DK$ events and $N_{D\pi}$ is 
   the number of $B \to D\pi$ events with the bachelor
   $\pi$ misidentified as a $K$.  The ratio $N_{DK}/N_{D\pi}$ is obtained
   by fitting the $\Delta E$ distribution for $B \to DK$ candidate events
   with $\mes > 5.27$ GeV (see Fig.~\ref{fig:mes_de}d).
   This is modeled as the sum of a combinatoric background
   function, a double-Gaussian for the $B\to D\pi$ background,
   and a Gaussian for the $B\to DK$ signal.  The parameters of the
   Gaussians in the $\Delta E$ fit are constrained from fits to the
   $\Delta E$ distributions of well-identified $B \to D\pi$ events
   with the bachelor $\pi$ assumed to be a $\pi$ or a $K$.

   \begin{figure}[htb]
      \includegraphics[clip,width=\linewidth]{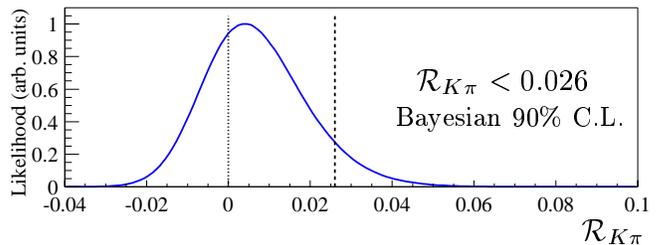}
   \caption{Likelihood as a function of ${\cal R}_{K\pi}$.  
   The integral 
   for $0<{\cal R}_{K\pi}< 0.026$
   is 90\% of the integral for ${\cal R}_{K\pi}>0$.  }
   \label{fig:likelihood_curve}
   \end{figure}

   We find ${\cal R}_{K\pi} = (4 \pm 12) \times 10^{-3}$,
   consistent with zero.  The number of signal, normalization,
   and peaking background events are
   $N_{sig} = 1.1 \pm 3.0$, $N_{DK} = 261 \pm 22$,
   $N_{peak}^D = r_D^2 N_{D\pi} = 0.38 \pm 0.07$, and
   $N_{peak}^{hhh} = 0.4 \pm 1.1$.
   The uncertainties are mostly statistical.
   From the likelihood, we set a Bayesian limit
   ${\cal R}_{K \pi} < 0.026$ at the 90\% confidence level (C.L.), assuming a
   constant prior probability for  ${\cal R}_{K\pi} > 0$
   (see Fig.~\ref{fig:likelihood_curve}).

   \begin{figure}[htb]
       \includegraphics[clip,width=\linewidth]{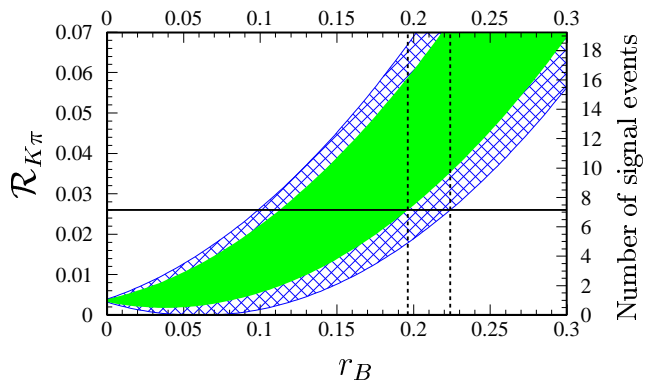}
   \caption{Expectations for ${\cal R}_{K\pi}$ and $N_{sig}$
   {\it vs.} $r_B$.  Filled-in area: allowed region for any value of 
   $\delta$, with a $\pm 1\sigma$ variation on $r_D$,
   and $48^{\circ} < \gamma < 73^{\circ}$.  Hatched area: additional
   allowed region with no constraint on $\gamma$.
   The horizontal line represents the 90\% C.L.  limit
   ${\cal R}_{K\pi} < 0.026$.  The dashed
   lines are drawn at $r_B = 0.196$ and $r_B = 0.224$. 
   They represent the 90\% C.L. upper limits on $r_B$ 
   with and without the constraint on $\gamma$.}
   \label{fig:ads_rate2}
   \end{figure}

   In Fig.~\ref{fig:ads_rate2} we show the dependence of
   ${\cal R}_{K\pi}$ on $r_B$, together with our limit.
   This is shown allowing a $\pm 1\sigma$ variation on $r_D$,
   for the full range $0^{\circ}-180^{\circ}$ for $\gamma$ and
   $\delta$, as well as with the restriction
   $48^{\circ} < \gamma < 73^{\circ}$ suggested by global CKM
   fits~\cite{ckmfitter}.  The least restrictive limit on $r_B$ 
   is computed assuming maximal destructive interference:
   $\gamma=0^{\circ}, \delta = 180^{\circ}$ or 
   $\gamma=180^{\circ}, \delta = 0^{\circ}$.  
   This limit is $r_B < 0.22$ at 90\% C.L.

   In summary, we find no evidence for 
   $B^{\pm} \to [K^{\mp}\pi^{\pm}]_D K^{\pm}$.  We 
   set a 90\% C.L. limit on the ratio ${\cal R}_{K\pi}$
   of rates for this mode and the favored mode
   $B^{\pm} \to [K^{\pm}\pi^{\mp}]_D K^{\pm}$.  Our limit is 
   ${\cal R}_{K\pi} < 0.026$ at 
   90\% C.L.
   With the most conservative assumption on the
   values of $\gamma$ and of the strong phases
   in the $B$ and $D$ decays, this 
   results in
   a limit on the ratio of the magnitudes
   of the $B^- \to \Dzb K^-$ and
   $B^- \to D^0 K^-$ amplitudes $r_B < 0.22$ at 90\% C.L.
   Our analysis suggests that $r_B$ is smaller than the
   value reported by the Belle collaboration, 
   $r_B = 0.26^{+0.11}_{-0.15}$~\cite{belleRb}, but 
   given the uncertainties the
   two results are not in disagreement.
   A small value of $r_B$ will make it difficult
   to measure $\gamma$ with 
   other
   methods~\cite{dk1}\cite{soffer}
   based on
   $B\rightarrow \dbarpnozero K$.

We are grateful for the excellent luminosity and machine conditions
provided by our \pep2\ colleagues, 
and for the substantial dedicated effort from
the computing organizations that support \babar.
The collaborating institutions wish to thank 
SLAC for its support and kind hospitality. 
This work is supported by
DOE
and NSF (USA),
NSERC (Canada),
IHEP (China),
CEA and
CNRS-IN2P3
(France),
BMBF and DFG
(Germany),
INFN (Italy),
FOM (The Netherlands),
NFR (Norway),
MIST (Russia), and
PPARC (United Kingdom). 
Individuals have received support from the 
A.~P.~Sloan Foundation, 
Research Corporation,
and Alexander von Humboldt Foundation.

   \end{document}